\documentclass[aps,prl,twocolumn,showpacs,superscriptaddress]{revtex4-1}  % for review and submission

\usepackage{graphicx}  % needed for figures
\usepackage{dcolumn}   % needed for some tables
\usepackage{bm}        % for math
\usepackage{amssymb}   % for math
\usepackage{natbib}
\usepackage{appendix}
\usepackage{hyperref}
\usepackage{color}
\newcommand{\abs}[1]{\left\vert#1\right\vert}
\newcommand{\ket}[1]{\left\vert#1\right\rangle}

\newcommand{\footnoteremember}[2]{
  \footnote{#2}
  \newcounter{#1}
  \setcounter{#1}{\value{footnote}}
}
\newcommand{\footnoterecall}[1]{
  \footnotemark[\value{#1}]
}

\begin{document}

% the following line is for submission, including submission to the arXiv!!
%\hspace{5.2in} \mbox{Fermilab-Pub-04/xxx-E}

\title{The effect of light assisted collisions on matter wave coherence in superradiant Bose-Einstein condensates}

\author{N.~S.~Kampel}
\email{nir.kampel@nbi.dk}
\affiliation{Niels Bohr Institute, Danish Quantum Optics Center QUANTOP, Copenhagen University, Denmark}

\author{A.~Griesmaier}
\affiliation{Niels Bohr Institute, Danish Quantum Optics Center QUANTOP, Copenhagen University, Denmark}
\affiliation{Physikalisches Institut, Universit\"{a}t Stuttgart, Pfaffenwaldring 57, 70569 Stuttgart, Germany
}
\author{M.P.~Hornbak~Steenstrup}
\author{F.~Kaminski}
\author{E.~S.~Polzik}
\author{J.~H.~M\"{u}ller}
\email{muller@nbi.dk}

\affiliation{Niels Bohr Institute, Danish Quantum Optics Center QUANTOP, Copenhagen University, Denmark}

\date{\today}

\begin{abstract}
We investigate experimentally the effects of light assisted collisions on the coherence between momentum states in Bose-Einstein condensates. The onset of superradiant Rayleigh scattering serves as a sensitive monitor for matter wave coherence. A subtle interplay of binary and collective effects leads to a profound asymmetry between the two sides of the atomic resonance and provides far bigger coherence loss rates for a condensate bathed in blue detuned light than previously estimated. We present a simplified quantitative model containing the essential physics to explain our experimental data and point at a new experimental route to study strongly coupled light matter systems.
\end{abstract}

%\pacs{03.75.-b(Matter waves),42.50.Nn(Super-radiance),42.50.Gy(BEC quantum optics)}
\maketitle

Elastic Rayleigh scattering of photons from atoms in a Bose-Einstein condensate (BEC) creates long-lived ripples in
the density distribution of the atomic cloud. Bosonic stimulation leads to a positive feedback mechanism enhancing
the formation of a matter-wave grating which scatters photons coherently predominantly along the directions of high optical
depth. This directed Rayleigh scattering is well known as Rayleigh superradiance and has been studied in BECs extensively
in different geometries \cite{Inouye(99)_FirstExpSuperRad,Schneble(03)_ShortPulseSuperRad,BarGill(07)_StrongPulseSuperRad,Hilliard(08)_SuperRad}. The coupled dynamics of superradiant (SR) scattering with simultaneous build-up of recoiling matter-waves and light fields have been
successfully described using Maxwell-Schr{\"o}dinger equations as well as rate equations derived from those \cite{Ketterle(10)_ColectiveFenomeBEC}.

Recently, those established descriptions of SR scattering have been challenged by a new model predicting a peculiar asymmetry
in the dependence of the dynamics on detuning of the drive light \cite{Deng(10)_BlueSuperRadTheory}. Clear experimental
evidence of such an asymmetry, which is pronounced at high atomic density and vanishes at low density, is reported in \cite{Deng(11)_BlueSuperRadExp}. This has sparked an ongoing debate \cite{Ketterle(10)_SuperRadComment,*Deng(10)_SuperRadComment} and it appears that the mechanism leading to the different
dynamics for below resonance (red) and above resonance (blue) tuning of the incident light is not yet fully understood.

To address this open question we present here an experimental study of the threshold behavior of SR light
scattering for a wide range of detunings up to $35~\mathrm{GHz}$. The observed detuning dependence rules out previous explanation attempts based on the action of dipole forces\footnoteremember{1}{See Supplementary Material for details}. We offer a physically motivated explanation for the asymmetry in the \emph{threshold} behavior of SR scattering based on detuning dependent loss of matter-wave coherence. Resonant excitation of close pairs of atoms to excited state molecular potentials and subsequent spontaneous decay provides a source of frequency shifted photons, which for blue detuned drive light can be trapped inside the BEC for a long time. The different accessible molecular branches with continuous and discrete spectra together with the fact that the inelastically scattered photons are inevitably red shifted with respect to the excitation light lead to an intrinsic red-blue asymmetry in the loss rates. We compare the results of a simplified rate equation model including such loss to our experimental data and find satisfactory agreement. The underlying mechanism offers exciting new possibilities to study optical excitations trapped deep inside dense ultracold atomic gases.

In our experiments, SR Rayleigh scattering is induced in a trapped BEC by illuminating it with an off-resonant light pulse
along the long axis of the condensate, which leads to recoiling atoms gaining two photon momenta. We use a $100~\mathrm{\mu s}$ rectangle
pulse and vary the pulse intensity to explore the onset of SR. BEC's of $^{87}Rb$ atoms in the $\ket{F=1, m_F=-1}$ hyperfine
state are prepared by evaporative cooling in a Ioffe-Pritchard magnetic trap \cite{Hilliard(08)_SuperRad}. We obtain prolate condensates containing $2.2\times 10^5$ atoms with in-trap Thomas-Fermi radii of $r_{\bot}\approx 6~\mathrm{\mu m}$ and $r_{\|}\approx60~\mathrm{\mu m}$ in the radial and axial directions, with no discernible thermal fraction. The pump light is detuned from the $\ket{F=1,m_F=-1}
\rightarrow \ket{F'=2,m_{F'}=-2}$ transition on the $D_1$ line of $^{87}Rb$ at $795~\mathrm{nm}$ and is circularly polarized. We measure the drive detuning ($\Delta=-2.57 \rightarrow 35~\mathrm{GHz}$) using a wavemeter ($\sim300~\mathrm{MHz}$ resolution) and a Fabry-Perot resonator ($1.5~\mathrm{GHz}$  FSR) referenced to a laser stabilized to saturated absorption features of the $^{85}Rb$ $D_1$ line. The pump beam is focused to a waist radius of $20~\mathrm{\mu m}$ (at $e^{-2}$) on the atoms. The pump pulse duration is chosen long enough to suppress backward (Kapitza-Dirac) scattering of atoms and short enough to neglect decoherence due to decaying overlap of matter wave packets in the later theoretical modeling. The range of explored pump detunings is chosen such as to have negligible pump light depletion at low detunings and is limited by our available laser power at high detunings.

\begin{figure} [t]
\centering
\includegraphics [width=0.75\textwidth,viewport= 125 250 700 545] {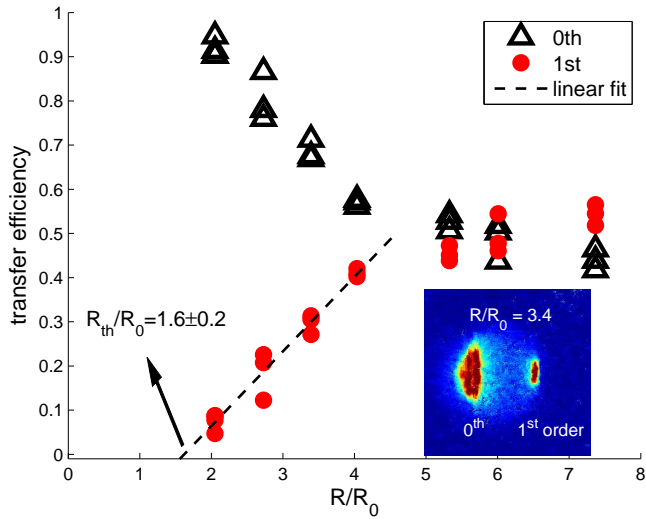}
\caption{(color online) The transfer efficiency as a function of normalized single atom Rayleigh scattering rate at $\Delta=5.24~\mathrm{GHz}$; Open triangles: population of the $0\hbar k$ momentum mode; Filled circles: population of the $2\hbar k$ momentum mode; Dashed line: linear fit; Inset: time of flight image of the atomic momentum distribution.} \label{fig:_extract threshold}
\end{figure}

The populations of atomic momentum modes ($0\hbar k$ \& $2\hbar k$) are extracted from absorption images taken after time-of-flight. At each detuning we measure the population transfer for different single atom Rayleigh scattering rates $R$, which
in simple models determine the timescale for the dynamics. We fit a straight line to the results with low scattering rate to
extract a phenomenological threshold pump rate $R_{th}(\Delta)$ where superradiant gain exceeds linear losses enough to start
significant population transfer during the interaction time. Figure~\ref{fig:_extract threshold} shows an example of transfer efficiency measurements at a blue detuning $\Delta=5.24~\mathrm{GHz}$. The scattering rate is normalized to the measured threshold rate $R_{0}$ at a red detuning of $\Delta=-2.57~\mathrm{GHz}$ \footnote{For red detuning above $2~\mathrm{GHz}$ SR dynamics has been shown previously to be independent of detuning \cite{Hilliard(08)_SuperRad}.}. A detuning asymmetry in the threshold and a saturation of the transfer efficiency at high scattering rates is visible.

\begin{figure} [tp]%[htp]
\centering
\includegraphics [width=0.75\textwidth,viewport= 100 250 700 550] {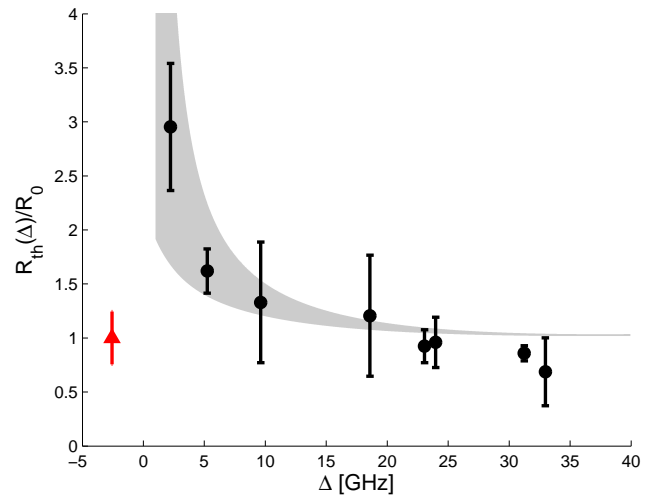}
\caption{(Color online) Normalized threshold scattering rate vs. detuning. Filled symbols show the measured threshold increase; Error bars designate $95\%$ confidence level; Gray shaded region depicts the expected threshold due to close range dipole-dipole interaction allowing for a factor of two variation of the light induced loss rate $L_{ge}$} \label{fig:_exp data}
\end{figure}

In Fig.~\ref{fig:_exp data} we present measured threshold rates as a function of pump laser detuning. A threshold increase up
to a factor of three is evident for low values of the blue detuning. Also shown in the figure is the expected threshold increase due to light assisted collisions calculated from the model presented in the following.

The starting point for the description of the coherent part of the SR process are coupled Maxwell-Schr\"{o}dinger equations \cite{Zobay(05)_MatterWaveSRtime,*Zobay(06)_MatterWaveSRspatial}. From these, we derive a rate equation for the number of atoms $N_2$ appearing in the recoil mode  $\dot{N}_2 = G\cdot(N_{2}+1)$, to describe the early stages of the dynamics where depletion of condensate atoms $N_0$ and pump light is not important\footnoterecall{1}. The rate constant for growth $G = R b_0$ depends for a fixed sample geometry linearly on the single atom Rayleigh scattering rate $R$ and on the effective resonant optical depth $b_0$ of the sample along the propagation direction of the superradiant light mode \cite{Ketterle(10)_ColectiveFenomeBEC}. Since the Rayleigh scattering rate varies symmetrically with laser detuning no asymmetry is predicted by this model\footnoteremember{2}{We arrive in the suppl. mat. at the same conclusion also for a 1-D model containing propagation effects.}. At this level of description Rayleigh superradiance does not have a threshold pump rate. To have a more realistic model of the onset, damping mechanisms for the coherence gratings need to be accounted for. We include loss rates $L_R=R$ to describe removal of $N_2$ atoms by spontaneous Rayleigh scattering, $L_{gg}(n_0)$ to describe damping by incoherent ground state collisions between $(N_0, N_2)$ pairs\footnoteremember{3}{A loss term describing coherence loss due to wavepacket separation can be freely added to this term, but is smaller than the collision term for our axial excitation geometry.}, and a loss rate $L_{ge}$ to account for light assisted collisions followed by radiation trapping. We also add a nonlinear loss rate $L_{nl}(N_0,\Delta)$, without discussing here the physical origin of possible nonlinear loss terms further\footnoterecall{1}, to account for processes that depend nonlinearly on the population of the superradiant modes as those suggested in \cite{Deng(11)_BlueSuperRadExp}. The resulting rate equation
\begin{equation}
\dot{N}_2 = G (N_2+1) -(L_R + L_{gg} + L_{ge})N_2 - L_{nl} N_2^2
\label{eq: rateEq}
\end{equation}
shows initial exponential growth when gain exceeds linear losses, i.e. $G>L_R+L_{gg}+L_{ge}$. Equality of gain and linear losses defines the threshold for SR scattering in this model. The threshold does not depend on $L_{nl}$. Similar to a depletion term, $L_{nl}$ clamps the growth rate later during the evolution when the population in the recoil mode becomes significant. We denote the threshold gain (Rayleigh rate) in the absence of the $L_{ge}$ term as $G_0$ ($R_0$) and parametrize the light induced loss rate as $L_{ge} = R \chi n_0 (1 + F \bar{n})$. Here, $R \chi n_0$ is the light assisted collision rate, with $n_0$ the condensate density and $\chi$ a molecular parameter, $F$ is the fraction of resonant photons produced in a collision, and $\bar{n}$ is the average number of subsequent scattering events for a resonant photon inside the cloud.

This parametrization of $L_{ge}$ puts emphasis on the role of $N_2$ atoms for the contrast and spatial coherence of the matter wave grating responsible for the amplified directed SR scattering. The matter wave grating amplitude can decrease or dephase either by direct participation and subsequent loss of an $N_2$ atom in a binary light assisted collision or by interaction of an $N_2$ atom with a resonant photon produced in any light assisted collision within the cloud. As we will show later, this second mechanism is far more important than direct loss of $N_2$ atoms\footnoteremember{4}{In the experiment we do not observe pronounced loss of atoms at any of the tested pump detuning values.}. Since in a fully quantum mechanical picture of SR scattering recoiling atoms and backscattered photons are created as correlated pairs contributing on equal footing to the gain, it is interesting to ask how important the loss of photon coherence due to inelastic collisions is for the net reduction of SR gain. The vast majority of binary collisions, which are the source for nearly isotropic incoherent resonant radiation, happens between $N_0$ atom pairs assisted by pump light photons. As long as depletion of the pump light is negligible, the direct influence of the frequency shifted radiation on the coherence of the light grating is marginal. It is the strong response of atoms to even minute amounts of resonant light which spoils the coherence of the matter wave and this way also the mutual coherence between light and matter waves.

Using Eq.~(\ref{eq: rateEq}) and the above expression for the loss term $L_{ge}$, the expected change of the threshold pump rate in the presence of light assisted collisions can be expressed as
\begin{equation}
\frac{R_{th}(\Delta)}{R_0} = \left(1 -\frac{\chi n_0 (1 + F \bar{n})}{b_{0}-1}\right)^{-1} ,
\label{eq: threshold}
\end{equation}
which is the quantity determined in our experiment. Due to the strong dependence of $\bar{n}$ on optical thickness, discussed later, the threshold increases markedly for dense and optically thick clouds, while it is unaltered in the limit of low density and optically dilute clouds, in accordance with the experimental observations in \cite{Deng(11)_BlueSuperRadExp}.

\begin{figure} %[htp]
\centering
\includegraphics [width=0.7\textwidth,viewport= 100 250 700 550] {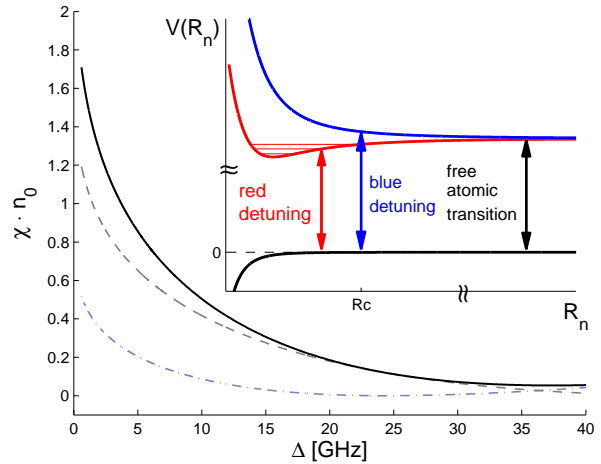}
\caption{(Color online) Ratio $\chi n_0$ of pair collision rate to Rayleigh scattering rate for an atomic density of $10^{14}~\mathrm{cm^{-3}}$ as a
function of detuning. Dashed and dash-dotted lines: Separate contributions from the $0_g^+$ and $1_u$ molecular potentials; Solid line: total ratio. Inset: Typical ground and excited state potentials for cold collisions in light fields in the detuning regime where the excited state structure is dominated by resonant dipole-dipole interaction. Red detuned light excites pairs to bound states (free-bound transition), while blue detuned light excites to a continuum (free-free transition).}
\label{fig:_loss rate}
\end{figure}

To allow for a comparison between model predictions and our experimental data, we turn now to a more detailed discussion of the three-step process leading to the coherence loss rate $L_{ge}$  and give quantitative estimates of the microscopic parameters $\chi$, $F$ and $\bar{n}$. Modeling off-resonant light scattering as purely elastic ceases to be a good approximation at high atom densities when the probability to excite close pairs of atoms becomes significant. To include this into the description a molecular point of view is necessary. Here, photons can be scattered with a significant frequency shift with a concomitant change in kinetic energy of the outgoing pair of atoms, high enough for both atoms to leave the trap. Since the outgoing photon has a frequency close to the atomic resonance it will scatter repeatedly inside the cloud before leaving. The dependence of $L_{ge}$ on detuning is inherently asymmetric since for red detuning only discrete bound molecular states can be resonantly excited (photoassociation resonances) while for blue detuning a continuum of states on repulsive molecular potentials is accessible (radiative heating) as sketched in the inset of Fig.~\ref{fig:_loss rate}. The initial light assisted collision step in the three-step process has
been used in the past to assess light induced atom loss rates from condensates \cite{Burnett(96)_moleLosses}. To estimate
quantitatively the event rate coefficient for binary light assisted collisions $R \chi$ appearing in $L_{ge}$ we switch
to a microscopic description of the collision employing the methods outlined in \cite{Julienne(96)_lightBinaryCollisionLosses}. Using a reflection approximation the event rate coefficient $R \chi$ is written as:
\begin{eqnarray}
R \chi = \frac{\pi \hbar}{\mu k_{\infty}} \times 4\pi^2 V_c^2 \times \frac{1}{D_C}\abs{\Psi_g (R_c,E)}^2
\end{eqnarray}

Here, $\mu$ is the reduced mass for a $^{87}Rb$ atom pair, $\hbar k_{\infty}$ is the relative momentum in the entrance channel
with corresponding kinetic energy $E$, and $R_c$ is the Condon radius where the molecule tunes into resonance. The Franck-Condon factor $\abs{\Psi_g (R_c,E)}^2 / D_C$, with $\Psi_g (R_c,E)$ the ground state scattering wave function and $D_c$ the difference in potential slopes, regulates the detuning dependence. The radiative coupling potential $V_c = V_{eg}(R_n) = b_C(R_n)\hbar\Omega_A$, where $\Omega_A$ is the atomic Rabi frequency and $b_C(R_n)$ a molecular parameter reflecting the change of the electronic wavefunction with internuclear distance $R_n$, varies only little in the range of atomic distances relevant here. The ground state
scattering wave function is calculated numerically by integrating the Milne equation. The accessible repulsive excited state molecular potentials are parameterized with dispersion coefficients $C_3(0_g^+) = 11.9~\mathrm{a.u.}$ and $C_3(1_u)=5.89~\mathrm{a.u.}$ following \cite{Movre(77)_EstimatingMolecPote1,*Movre(80)_EstimatingMolecPote2}. The main panel of Fig.~\ref{fig:_loss rate} shows the ratio
$\chi n_0$ between light assisted collision rate and isolated atom Rayleigh scattering rate at a typical BEC density of $n_0=10^{14}~\mathrm{cm^{-3}}$ for excitation to the different molecular potentials.

To model the second step, the frequency shift of pump photons assisting a collision, we estimate the frequency spectrum of light exiting the collision complex and determine the fraction $F$ emitted within one natural linewidth of the bare atomic resonance. Using a semiclassical wave packet approach we calculate the trajectory of an excited atom pair separating from the Condon radius and use energy conservation to determine the frequency shift of light emitted along the way\footnoterecall{1}. Due to the high acceleration on the repulsive molecular potential almost all of the atom pairs reach the asymptotic kinetic energy before decaying, making the frequency redistribution function sharply peaked around the atomic resonance. Taking into account hyperfine branching in the decay around $43\%$ of the emitted light is fully resonant with the hosting cloud for all $\Delta > 0$ used in the experiment.

As the third ingredient to assess the damping of the matter wave coherence we need to find the average number of scattering
events $\bar{n}$ for resonant photons before leaving the cloud. Neglecting further frequency redistribution we use a simplified
Holstein model, essentially a diffusive transport equation for light intensity in a medium of high optical depth, to describe
radiation trapping \cite{Fioretti(98)_radiationTrapping}. We use the decay time of the slowest Holstein mode $\tau_0^{el} = \gamma b^2 \tau_{nat} \simeq \bar{n} \tau_{nat}$ to calculate $\bar{n}$ \cite{Labeyrie(03)DiffusionColdCloud}. Here, $\gamma$ is a geometry parameter, $\tau_{nat}$ is the radiative lifetime of the excited atomic state, and $b$ is the optical depth. For a simple estimate we assume a Gaussian spherical geometry ($\gamma \simeq 0.06$) with an optical depth equivalent to the geometric mean along the different condensate axes. For our parameters we find $\bar{n} > 1000$.

The gray area in Fig.~\ref{fig:_exp data} depicts the calculated threshold increase via Eq.~(\ref{eq: threshold}) for the
experiment. Given the several rather crude assumptions in the calculation of the microscopic model parameters, together with smaller systematic uncertainties in the experimental parameters atom number and density, we allow for a factor of $2$ variation of the calculated loss rate $L_{ge}$ in Fig.~\ref{fig:_exp data}. We note that the exact spectral signature is sensitive to
molecular hyperfine structure, which is not taken into account in our model potentials. While the qualitative and near quantitative agreement between data and prediction is satisfying to see, the rate equation and radiation trapping model applied does not do full justice to the underlying complicated many-body physics. Simple inspection of the threshold condition reveals that at our highest observed threshold increase more than $60\%$ of the atoms should have interacted with a trapped resonant photon. The little observed recoil heating in the experiment is clearly incompatible with a picture of individual atoms receiving random recoil kicks from an isotropic radiation field. In fact, at resonance and high density photonic and atomic degrees of freedom mix strongly, forming polariton type excitations with an effective mass very different from the bare atomic mass \cite{Shlyapnikov(90)}. In using the radiation diffusion model we implicitly assume that the dephasing rate is still governed by the bare atomic decay rate $\Gamma$.

Whatever the precise nature of the trapped excitations is, their incoherent production by collisions and slow diffusion
implies the presence of electronic excitation inside the cloud many natural lifetimes after the pump light has left the cloud. Photoionization out of the excited state can provide a critical test of the model but also a tool to study the temporal dynamics of the polaritons in detail. The spatial structure of phase damage can be mapped out by matter wave shearing interferometry \cite{Simsarian(00)_phase_evolving_BEC_image}. In future work, it will be interesting to study the different response of dense clouds to resonant light applied from the outside or created directly inside the medium  \cite{Politzer(91)_incidentLightBEC}.

In conclusion, we have studied experimentally the threshold asymmetry of Rayleigh superradiance. We have developed a simple, yet quantitative, model to explain our data and discussed the underlying physics, which can hopefully serve as a useful guide for more rigorous theoretical studies. The mechanism for deposition of resonant photons deep inside a dense cloud might provide a promising route to observe Anderson localization of light in cold atoms \cite{Akkermans(08)_photon_localization_superradiance}.

We acknowledge support by the EU through FP6 RTN EMALI and through the projects HIDEAS and Q-ESSENCE. We thank A. S. S{\o}rensen for inspiring discussions.

\newpage 
\onecolumngrid

\section{Supplementary Material}

\subsection{Derivation of the rate equation}

In the main text we use a simplified rate equation for the number of recoiling atoms to describe the onset of superradiance (SR) in the absence of incoherent losses. Here, we sketch the steps to arrive at the rate equation, which averages over propagation effects, starting from Maxwell-Schr{\"o}dinger equations, which include all propagation effects. The basic coupled evolution equations for light and matter fields have been presented in the literature already several times \cite{Zobay(05)_MatterWaveSRtime,*Zobay(06)_MatterWaveSRspatial,Hilliard(08)_SuperRad,BarGill(07)_StrongPulseSuperRad}. On the way to the rate equation we discuss, in particular, the various approximations that enter the derivation, also in view of recent attempts to explain the detuning asymmetry in superradiance based on approximate analytic solutions of the Maxwell-Schr{\"o}dinger equations and modifications thereof \cite{Deng(10)_BlueSuperRadTheory,Deng(11)_BlueSuperRadExp}.

The starting point are the mean-field Gross-Pitaevski equation for the matter field $\psi$ in the electronic ground level including an effective coupling term to the light field and a classical wave equation for the propagating electric field $\textbf{E}$ with a polarization term to describe the coherent radiation by the driven atoms:
\begin{eqnarray}
i\hbar \frac{\partial}{\partial t} \psi = -\frac{\hbar^2}{2M}\nabla^2\psi + \frac{\left(\textbf{d}^+\cdot \textbf{E}^-\right)\left(\textbf{d}^- \cdot \textbf{E}^+\right)}{\hbar\Delta} \psi  + g_0\abs{\psi}^2\psi \label{eq:schrodinger}\\
\left( \nabla^2 - \frac{1}{c^2} \frac{\partial^2}{\partial t^2} \right)\textbf{E}^{\pm} = \frac{1}{c^2 \epsilon_0} \frac{\partial^2}{\partial t^2} \textbf{P}^{\pm} . \label{eq:maxwell wave}
\end{eqnarray}
Here, superscripts $\pm$ denote positive and negative frequency components, $M$ is the atomic mass for $^{87}Rb$, $\hbar$ is the reduced Planck constant, $\textbf{d}$ is the atomic dipole matrix element on the driven electronic transition, $c$ is the speed of light, and $\epsilon_0$ is the free space permittivity. The nonlinear term proportional to $g_0$ describes the mean field energy from ground state van der Waals interaction. The macroscopic polarization is given by $\textbf{P}^{\pm} = -\left|\psi\right|^2
\textbf{d}\left(\textbf{d} \cdot\textbf{E}^{\pm}\right)/\left(\hbar\Delta\right)$. Several approximations have already been applied to arrive at this form of the equations. The trapping potential for the atoms has been dropped, since it has negligible influence on the dynamics on the time scale of the interaction with the pump pulse. Implicitly the effect of the trapping potential is, of course, contained in the initial density distribution of the cloud. Only the Rayleigh scattering channel back into the initial Zeeman sublevel is considered for the superradiant dynamics, since the inhomogeneous magnetic field destroys rapidly the coherence between different sublevels. More importantly, the dipole response of the atoms is calculated in second order perturbation theory for an isolated atom in a rotating wave and low saturation approximation to eliminate excited states adiabatically. The radiative damping term $i\Gamma/2$, which formally needs to be added to the detuning $\Delta$, has negligible influence for the range of detunings considered later on, and is thus left out. Doing the adiabatic elimination at the single atom level, the light mediated interaction between atoms in their near-field and its impact on the scattering properties are neglected. We choose to take these effects into account later by switching to a molecular picture for close pairs of atoms.  Effective macroscopic descriptions including the near-field interaction, derived several times in the literature, have many subtleties \cite{Sokolov(09)_denseScattererGas,Lewenstein(96)_QscatteringWeakCWboson}. In particular, frequency redistribution processes and radiation trapping, invoked as the source of decoherence in the main text, are buried deep under the formalism and are only hard to recognize in the macroscopic treatment, which concentrates on the stationary linear response of the scattering medium.

At this stage, as long as the Maxwell-Schr{\"o}dinger equations are solved simultaneously, all coherent light mediated interaction between driven dipoles in their far-field and all dipole forces are still fully accounted for.
Keeping the full 3-D description throughout is cumbersome, so the next step is a reduction to 1-D. Two potentially important effects, depending on the specific geometry of an experiment, that are lost in a 1-D description are the transverse components of the dipole force exerted by the light field distribution on the atoms and the corresponding back action on the light (diffraction and lensing). In our experiments, the transverse dipole force accelerates the atoms radially outward, increasingly for higher blue detunings. At the threshold for SR scattering the peak density is calculated to drop, during the short interaction time, by about $3\%$ for the lowest blue detuning, while for the highest blue detunings a density change of about $20\%$ is expected. This systematic effect is not taken into account in the data analysis and model presented in the main text, but can possibly explain why the observed threshold rates at high blue detunings dip below the red detuning reference value, via a reduced rate of incoherent ground state collisions. Lensing of pump light due to the spatial variation of the refractive index, modifies the intensity distribution and hence the effective light-atom coupling. This effect can mimic the observed detuning dependence of the threshold pump rate. For the lowest blue detuning in our experiment, where the effect is biggest, we estimate, from numerical 3-D simulations of light intensity inside the sample, a relative change of the Rayleigh scattering rate of 20\% with respect to the red detuning reference value, thus significantly smaller than needed to explain the experimental observations by lensing alone. We have not included this systematic effect in the data presented in the main text.

This said, we simplify the above equations by going to an effective 1-D geometry assuming a constant transverse cross section $A$ of the interaction region. This keeps longitudinal components of interaction and dipole forces only.
To simplify the equations further we split the electric field into forward and backward propagating modes and introduce slowly varying envelope functions for the modes denoted by subscripts $\pm$ in the following. Likewise, the matter wave function is split into recoil modes with slowly varying envelopes as $\psi(z,t) = \sum_{m=2n} \psi_{m}(z,t) e^{-i\left( \omega_m t - m k z\right)}$ where $m=2n$ and $n$ is an integer number (SR order number), $\omega_m=m^2\omega_r$, $\omega_r=\frac{\hbar k^2}{2M}$ is the recoil frequency, and $k$ is the wave number. Since the condensates used in the experiments are much longer than an optical wavelength ($L\simeq 100\lambda$) the mode functions are orthogonal to a very good approximation. As the next step we transform the equations to dimensionless form, by defining electric field, time and length units via $\textbf{E}_{\pm} = \varepsilon_{\pm} \sqrt{\frac{\hbar\omega}{2\epsilon_0}\cdot \frac{2\omega_r}{c A}}$, $\tau=2\omega_r t$ and $\xi=kz$. Using the slowly varying envelope approximation the transformed equations read:
\begin{eqnarray}
\frac{\partial\varepsilon_+}{\partial \xi} = -i\Lambda \sum_m \left\{ \varepsilon_+\abs{\psi_m}^2 + \varepsilon_-
\psi_{m-2}^*\psi_m e^{-2i(m-1)\tau} \right\} \label{eq: ep prop} \\
\frac{\partial\varepsilon_-}{\partial \xi} = i\Lambda \sum_m \left\{ \varepsilon_+\psi_{m+2}^*\psi_m e^{2i(m+1)\tau} +
\varepsilon_-\abs{\psi_m}^2 \right\} \label{eq: em prop} \\
\frac{\partial\psi_{m}}{\partial \tau} = \frac{i}{2} \frac{\partial^2\psi_{m}}{\partial \xi^2} - m
\frac{\partial\psi_{m}}{\partial \xi} -i\Lambda \left( \abs{\varepsilon_+}^2 + \abs{\varepsilon_-}^2 \right)
\psi_m -i\frac{\omega_{MF}}{2\omega_{r}}\sum_n \sum_{l}\psi_n^*\psi_{n-l}\psi_{m+l} e^{-il\left(m-n+l\right)\tau} \nonumber \\
-i\Lambda \varepsilon_+^* \varepsilon_- e^{-2i(m+1)\tau} \psi_{m+2} -i\Lambda \varepsilon_-^* \varepsilon_+ e^{2i(m-1)\tau}\psi_{m-2} . \label{eq:pm with kin}
\end{eqnarray}
The coupling constant is expressed as $\Lambda = \Gamma \sigma_0/\left(4\Delta A\right)$ and $\omega_{MF}$ denotes the mean-field ground state interaction. The ratio of atomic absorption cross-section $\sigma_0$ to sample cross section $A$ and the detuning $\Delta$ in units of the linewidth $\Gamma$ determine the strength of the effective atom-light interaction. The first two terms on the r.h.s. of Eq.~(\ref{eq:pm with kin}) describe wave packet spreading and recoil induced drift of the matter wave envelopes, which ultimately leads to coherence loss by spatial separation. In our experiments the influence of these terms is small due to the short interaction time and long sample length, hence we drop them in the following.

To connect the equations more directly to observable quantities we switch to density matrix elements instead of mode amplitudes.  Since we are interested mainly in the onset of superradiance we restrict the number of recoil modes for the matter wave to the first two. This makes the system formally equivalent to a coherent two-level amplifier/absorber, with a weak nonlinear contribution due to the mean field interaction.
\begin{eqnarray}
\frac{\partial \abs{\psi_0}^2}{\partial \tau} = i\Lambda \left(\varepsilon_-^*\varepsilon_+ e^{2i\tau} \psi_0\psi_{2}^*
-\varepsilon_+^* \varepsilon_- e^{-2i\tau} \psi_0^*\psi_{2}\right) \label{eq: p02 prop} \\
\frac{\partial \abs{\psi_2}^2}{\partial \tau} = -i\Lambda \left(\varepsilon_-^* \varepsilon_+ e^{2i\tau} \psi_2^*\psi_{0}
- \varepsilon_+^* \varepsilon_- e^{-2i\tau}\psi_2\psi_{0}^* \right) \label{eq: p22 prop}\\
\frac{\partial}{\partial \tau}\psi_0^*\psi_{2} = i\Lambda \varepsilon_-^* \varepsilon_+ e^{2i\tau} \left( \abs{\psi_{2}}^2 - \abs{\psi_{0}}^2 \right) + i\frac{\omega_{MF}}{2\omega_{r}} \left(\abs{\psi_{2}}^2 - \abs{\psi_0}^2 \right) \psi_0^* \psi_{2}  \label{eq: p0p2 prop}
\end{eqnarray}
The last term in Eq.~(\ref{eq: p0p2 prop}) stems from the extra energy cost to create a density modulation in the interacting cloud and describes e.g. the mean field shift of a Bragg resonance.
The light flux propagation equations read:
\begin{eqnarray}
\frac{\partial\abs{\varepsilon_+}^2}{\partial \xi} = i\Lambda \left( \psi_{2}^*\psi_{0} e^{2i\tau} \varepsilon_-^*\varepsilon_+ - \psi_{0}^*\psi_2 e^{-2i\tau} \varepsilon_+^*\varepsilon_- \right) \\
\frac{\partial\abs{\varepsilon_-}^2}{\partial \xi} = i\Lambda \left( \psi_{2}^*\psi_0 e^{2i\tau} \varepsilon_-^*\varepsilon_+ - \psi_0^*\psi_{2} e^{-2i\tau} \varepsilon_+^*\varepsilon_- \right) \\
\frac{\partial}{\partial \xi}\varepsilon_-^*\varepsilon_+ = - i\Lambda \psi_0^*\psi_{2}e^{-2i\tau} \left( \abs{\varepsilon_+}^2 + \abs{\varepsilon_-}^2 \right) - 2i\Lambda \left(\abs{\psi_0}^2 + \abs{\psi_2}^2\right)\varepsilon_-^* \varepsilon_+  .\label{eq: e-e+}
\end{eqnarray}
The last term in  Eq.~(\ref{eq: e-e+}) reflects the modification of light wavelength due to the refractive index of the cloud.
Writing the complex coherences in polar form as $\varepsilon_-^*\varepsilon_+ = \rho_l e^{i\phi_l}$ and $\psi_0^* \psi_{2} = \rho_a e^{i\phi_a}$ brings the equations into a form suitable for further discussion.
\begin{eqnarray}
\frac{\partial}{\partial \tau} |\psi_2|^2 = -\frac{\partial}{\partial \tau} |\psi_0|^2 =
2\Lambda\rho_l\rho_a \sin \left(2\tau+\phi_l-\phi_a \right) \label{eq: p2init}\\
\frac{\partial}{\partial \xi}|\varepsilon_-|^2 = \frac{\partial}{\partial \xi}|\varepsilon_+|^2 = -2\Lambda \rho_a\rho_l \sin\left( 2\tau + \phi_l-\phi_a\right) \label{eq: e init}\\
\frac{\partial}{\partial \xi}\rho_l = -\Lambda \left(|\varepsilon_+|^2 + |\varepsilon_-|^2\right) \rho_a \sin(2\tau + \phi_l - \phi_a) \label{eq: rho_l init}\\
\frac{\partial}{\partial \tau}\rho_a =\Lambda \left(|\psi_0|^2 - |\psi_2|^2 \right) \rho_l \sin\left( 2\tau + \phi_l-\phi_a\right)  \label{eq: rho_a init}
\end{eqnarray}
One can recognize the first equality in Eq.~(\ref{eq: p2init}) as the local conservation of atom number, which is a consequence of neglecting wavepacket drift and spread. Similarly, the first equality in Eq.~(\ref{eq: e init}) expresses a continuity equation for the photon density, a necessary consequence of the adiabatic elimination of excited atomic states.

We note that the terms describing the mean-field interaction and the refractive index drop from the magnitude and coherence equations. These terms will weakly influence the evolution of the phase of matter and light gratings ($\phi_a$ \& $\phi_l$). With our choice of the pump laser frequency as the carrier frequency for both forward and backward propagating light modes, the time dependence of $\phi_l$ acquires the recoil shift of the backscattered light and compensates the explicit time dependence in Eqs.~(\ref{eq: p2init}-\ref{eq: rho_a init}).

To model the onset of superradiant scattering we make use of the specific initial and boundary conditions in the experiments, i.e. $|\psi_0|^2 \gg |\psi_2|^2$ and $|\varepsilon_+|^2 \gg |\varepsilon_-|^2$. At first sight the equations seem to imply an
odd symmetry in the dependence on the sign of pump laser detuning ($\Lambda\propto\Delta^{-1}$), and hence to explain the observed asymmetry in the SR threshold. A closer inspection reveals, that this is not the case. Due to the sinusoidal dependence on the relative phase of light and matter wave gratings, Eqs.~(\ref{eq: p2init}-\ref{eq: rho_a init}) support runaway solutions, growing nearly exponentially in time and in space, for both signs of the parameter $\Lambda$ ($\Delta$). The SR is triggered by spontaneous Rayleigh scattering that creates random gratings and provides a seed for the growth \footnote{For analytic solutions of completely analogous quantum dynamics see e.g.\cite{Raymer(81)_raman_solutions,*Mishina(07)_RamanMemory}.}.

If boundary conditions are such that both $|\varepsilon_-|^2$ and $|\varepsilon_+|^2$ are strong light fields incident on the sample, the coupled equations describe just Bragg diffraction of matter waves in a (walking) standing wave including the backaction of atoms onto the light field. The different relative phase for gain between light and matter interference patterns for blue and red detuning can be understood easily in an optical lattice picture, when considering the spatial structure of Bloch waves at the band edge. Similarly, for initial conditions such that both atomic recoil modes are macroscopically populated, reflection of light from a density grating is described.

Returning to SR scattering, in a minimalistic approach the equations can be reduced to a zero dimensional system leading to the rate equation for the number of atoms appearing in the recoil mode. To do this we assume initial homogeneous matter wave coherence $\rho_a(\tau=0)$ over the sample corresponding to one delocalized atom in the recoil mode and perfect phase matching conditions $\sin(2\tau + \phi_l - \phi_a) \simeq 1$ which is valid in the early stages of the dynamics. Now, Eq.~(\ref{eq: rho_l init}) is solved subject to the boundary condition that $\rho_l(\xi_{max}) = 0$. The result is inserted into Eq.~(\ref{eq: p2init}), leading to:
\begin{equation}
\frac{\partial}{\partial \tau} |\psi_2|^2 =
2\Lambda^2\left|\epsilon_+\right|^2
\rho_a^2 \left(1 - \frac{\xi}{\xi_{max}}\right) .
\label{eq: p2rate}
\end{equation}
Integrating this over the length of the sample renders a rate equation for the number of atoms in the recoil mode as:
\begin{equation}
\frac{\partial}{\partial \tau} N_2 = \Lambda^2 N_0
\left|\epsilon_+\right|^2 (N_2 + 1) .
\label{eq: p2integral}
\end{equation}
Restoring physical units the gain constant appearing on the r.h.s. can be written now as $G = R b_0$, with the single atom Rayleigh rate $R$ and the on-resonance optical depth $b_0$ as stated in the main text.

We conclude this derivation with a brief discussion of basic scaling properties of the dipole force and point in this context at some fundamental flaws in the model put forward by Deng {\textit{et al.}} \cite{Deng(11)_BlueSuperRadExp} to explain the observed detuning asymmetry in SR scattering. In a low saturation approximation the dipole force can be written as the gradient of the polarization energy of the atoms. For the corresponding potential we have, thus, $U_{dip} \propto I/\Delta$. Here, $I$ is the {\emph{total}} intensity at the location of an atom. It is instructive to consider the detuning dependence of the dipole potential $\tilde{U}_{dip}$ for the case of constant Rayleigh scattering rate $R$, which is of relevance for the pump light in SR scattering experiments. Since $R \propto I/\Delta^2$, we must have $\tilde{U}_{dip} \propto \Delta$, i.e. the dipole force and potential increase in this case linearly with the detuning. Finally, for the onset of SR scattering the relevant dipole potential stems from the interference pattern between pump light at constant Rayleigh rate and scattered light. Here, the scattered light electric field $\tilde{E}_s$ is independent of detuning, while for the pump light field strength we have $\tilde{E}_p \propto \Delta$, which means that the strength of the dipole force is independent of detuning for this case. This reflects nicely the fact that in both coupled wave and rate equation models for SR the Rayleigh rate is the only relevant pump light parameter for the dynamics and that the very mechanism of SR Rayleigh scattering can be understood in terms of dipole forces. From these scaling arguments we conclude that the experimentally observed detuning dependence of the threshold asymmetry cannot be reconciled with models based solely on the action of dynamically evolving dipole forces.

In the model of Deng {\textit{et al.}} a dipole potential is invoked as being produced by a seed photon pulse, that is amplified upon propagation through the pump light dressed medium \footnote{For a fair comparison, the seed pulse flux should be replaced by the Rayleigh rate into relevant modes.}. This ignores both the intensity of the pump light, as well as the interference term between pump light and scattered light. Even in the most favorable case of plane wave pumping the dipole forces due to the interference term are orders of magnitude bigger than the contribution of the scattered light intensity alone. Due to this neglect Deng {\textit{et al.}} also arrive at an incorrect scaling with detuning. Deng {\textit{et al.}} suggest furthermore in their model, that the structure factor for light scattering is directly modified by their invoked incomplete dipole potential. Since a structure factor depends sensitively on interaction and correlation properties of the constituent particles, this suggestion appears entirely unfounded without a detailed look at microscopic properties.

\subsection{Frequency redistribution function}
To get the frequency spectrum of photons produced in binary collisions, we use the Ehrenfest theorem to calculate by classical mechanics the trajectory of the excited state wave packet in the repulsive molecular potential. Knowing the kinetic energy as a function of time along the trajectory, allows to transform the probability distribution for decay as a function of time to the  spectral distribution. The wave packet approach is justified by the same stationary phase argument that is used to calculate the Franck-Condon factor for the upward transition. The calculation is completely analogous to the survival probability estimate used in the Gallagher-Pritchard model for binary collisions in red detuned light fields \cite{Pritchard_model}.

We start by considering the total energy $E$ available in a generic two-body (half-) collision in a repulsive $r^{-3}$ potential,
\begin{eqnarray}
E = T + V \\
\hbar\Delta = \frac{\mu}{2} \dot{r}^2 + \frac{C_3}{r^3} , \label{eq: collision energy}
\end{eqnarray}
with $T$ and $\mu$ the kinetic energy and reduced mass, respectively. We do not consider a centrifugal potential term. The ground state scattering wave function has $s$-wave symmetry, while the electronic angular momentum coupling in the non-centrosymmetric dipole potential is accounted for by the designation of the molecular state, parametrized by $C_3$. The energy of accessible (l=0, l=1) rotational state continua differs for the smallest Condon radii by less than the atomic natural line width $\hbar\Gamma$ which is dwarfed by the total collision energy $\Delta\gg\Gamma$. The detuning is defined as $\Delta = \omega_L-\omega_0$, where $\omega_L$ is the laser frequency and $\omega_0$ is the atomic line resonance frequency.

\begin{figure} %[htp]
\centering
\includegraphics [width=0.75\textwidth,viewport=5 225 600 550,clip] {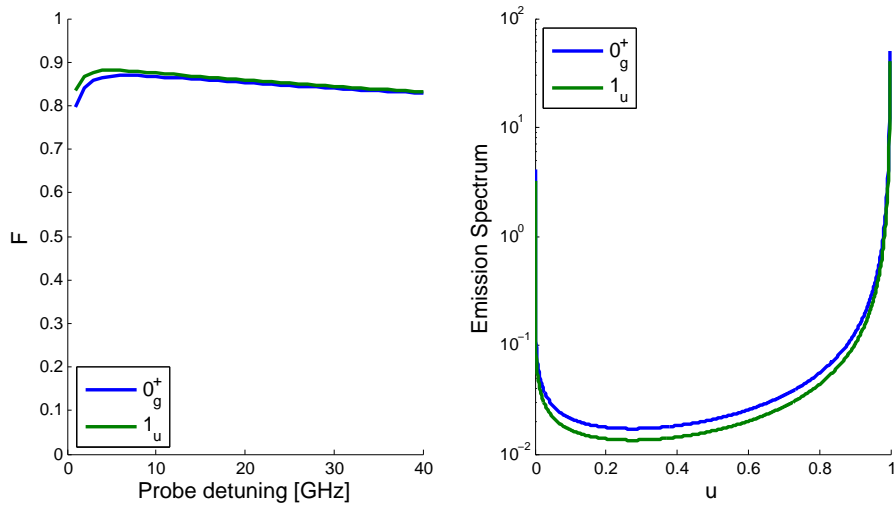}
\caption{(Color online) Left figure: Fraction of photons emitted in a frequency interval of width $\Gamma$ above the atomic resonance as a function of detuning. Right figure: emission spectrum ($\frac{dP}{du}$) as a function of $u=\left(\omega_l-\omega\right)/\Delta$, evaluated at $\Delta=3GHz$.}
\label{fig:_emissionSpec_F}
\end{figure}

Simple algebraic manipulation allow us to derive the differential equation describing the temporal change of kinetic energy along the trajectory as
\begin{equation}
\dot{T} = \left(\frac{18}{\mu C_3^{2/3}}\right)^{1/2} T^{1/2} (E-T)^{4/3} ,
\end{equation}
which can be integrated by separation of variables. Introducing scaled variables $u = T/E$ and $\tau = \Gamma_m t$, where
$1/\Gamma_m$ is the radiative lifetime of the excited molecular state, we write the solution as:
\begin{equation}
\tau(u) = \frac{\Gamma_m}{\alpha\Gamma} \int_0^u \frac{dx}{x^{1/2}(1-x)^{4/3}} \; .
\end{equation}
The coefficient $\alpha = 6(\Delta/\Gamma)^{5/6}(2\omega_{r}/\Gamma)^{1/2} (\hbar\Gamma/(C_3k^3))^{1/3}$ contains all physical parameters of the specific system, while the integral can be expressed in terms of hypergeometric functions. When evaluated  numerically, care must be taken to treat the singularities of the integrand correctly.

The probability density for decay to the electronic ground state of the colliding atom pair is given by
\begin{equation}
\frac{dP}{dt} = \Gamma_m \exp(-\Gamma_m t) .
\end{equation}
Moving to the scaled kinetic energy as the independent variable and transforming the differential accordingly we arrive at
\begin{equation}
\frac{dP}{du} = \frac{\Gamma_m}{\alpha\Gamma} \frac{1}{u^{1/2}(1-u)^{4/3}}
\exp\left[-\frac{\Gamma_m}{\alpha\Gamma}\int_0^u \frac{dx}{x^{1/2}(1-x)^{4/3}}\right] .
\end{equation}
Recognizing that the variable $u \in [0..1]$ maps the energy of the outgoing photon on the interval $[\omega_0 + \Delta ..\omega_0]$ the fraction $F$ of photons emitted in a frequency interval of width $\Gamma$ above the atomic resonance can be written as
\begin{equation}
F = \exp\left[- \frac{\Gamma_m}{\alpha\Gamma} \int_0^{1- \Gamma/\Delta} \frac{dx}{x^{1/2}(1-x)^{4/3}}\right] .
\end{equation}
The spectrum of fluorescence for excitation at $\Delta = 500\Gamma$ as well as the fraction $F$ as a function of detuning are shown in Fig.~\ref{fig:_emissionSpec_F}. The additional (small) broadening due to the finite emission time $\Gamma_m^{-1}$ is not taken into account in this simple calculation. We finish this calculation with the remark, that the recoil shift for the emitted radiation, which is, of course, negligibly small compared to the red shift compensating the change of relative kinetic energy, must be evaluated using the total mass of the composite radiating system, i.e. twice the atomic mass. Providing the answer to the equivalent questions about deposited energy and recoil for the case of resonant radiation incident on a whole group of close atoms, is an interesting but highly nontrivial task, in our view.

%\bibliographystyle{prsty}
%\bibliographystyle{plain}
%\bibliographystyle{apsrev4-1}
%\bibliography{../bibliographyFull}


\end{document}